\documentclass[12pt,preprint]{aastex}

\def\spose#1{\hbox to 0pt{#1\hss}}
\def\lta{\mathrel{\spose{\lower 3pt\hbox{$\mathchar"218$}}
     \raise 2.0pt\hbox{$\mathchar"13C$}}}
\def\gta{\mathrel{\spose{\lower 3pt\hbox{$\mathchar"218$}}
     \raise 2.0pt\hbox{$\mathchar"13E$}}}

\def\figure#1#2 {\par{\narrower\noindent {\bf Fig. #1}
   \hskip 2mm #2\par}\bigskip\noindent}
\def\table#1#2 {\par{\narrower\noindent {\bf Tab. #1}
   \hskip 2mm #2\par}\bigskip\noindent}

\shorttitle{A New Version of Reimers' law of Mass Loss}

\shortauthors{K.-P. Schr\"oder, M. Cuntz}

\begin{document}


\title{A New Version of Reimers' law of Mass Loss \\
Based on a Physical Approach}


\author{K.-P. Schr\"oder}
\affil{Astronomy Centre, University of Sussex, Brighton, BN1 9QH, England;}
\email{kps@central.susx.ac.uk}
\and
\author{M. Cuntz}
\affil{Department of Physics, University of Texas at Arlington, Box 19059,}
\affil{Arlington, TX 76019, USA;}
\email{cuntz@uta.edu}


% \altaffiltext{1}{XXXXXX}

\pagebreak


\begin{abstract}
We present a new semi-empirical relation for the mass loss of cool 
stellar winds, which so far has frequently been described
by ``Reimers' law''.  Originally, this relation was based solely on
dimensional scaling arguments without any physical interpretation.  
In our approach, the wind is assumed to result from the spill-over of the extended
chromosphere, possibly associated with the action of waves, especially
Alfv\'en waves, which are used as guidance in the derivation of the
new formula.  We obtain a relation
akin to the original Reimers law, but which includes two new factors.
They reflect how the chromospheric height depends on gravity and how the
mechanical energy flux depends, mainly, on effective temperature.
The new relation is tested and sensitively calibrated by modelling 
the blue end of the Horizontal Branch of globular clusters. 
The most significant difference from mass loss rates predicted by
the Reimers relation is an increase by up to a factor of 3 for 
luminous late-type (super-)giants, in good agreement with observations.
\end{abstract}



\clearpage


%%%%%%%%%%%%%%%%%%%%%%%%%%%%%%%%%%%%%%%%%%%%%%%%%%%%%%%%%%%%%%%%%%%%%%%%%%%%

\section{Introduction}

Empirical mass loss formulas are pivotal for the construction of
empirical and semiempirical stellar atmosphere and wind models,
stellar evolution computations and studies of
the interstellar medium, among other topics.
Historically, the mass loss rate $\dot{M}$ of late-type giants and
supergiants has been described by
``Reimers' law", given as $\dot{M} = \eta \cdot \frac{L_* R_*}{M_*}$
\citep{rei75,rei77}, with $L_*$, $R_*$, $M_*$
as stellar luminosity, radius, and mass, respectively, given
in solar units, and $\eta$ is a fitting parameter.
Other empirical mass loss formulas have been presented by
\cite{lam81}, \cite*{dej88}, and \cite{nie90}, but they do 
not distinguish between the strong, and now well-described dust-driven 
winds \cite[e.g.,][]{wac02}, and the physically very different case
of nondust-driven winds.

Despite its wide-ranging success, the mass loss formula by
\citeauthor{rei75} suffers from two important deficiencies.
First, it is solely based on dimensional scaling arguments
without any physical interpretation.  In particular, the appearance
of the stellar luminosity in the formula is awkward noting that for
cool star winds, with the exception of molecule-driven and
dust-driven winds, the luminosity of the star is not expected
to be relevant \cite[e.g.,][]{hol85}.  In fact, the Reimers law
seems to suggest that a certain fraction of the stellar luminosity
$L_*$ is utilized to lift the wind material from the photosphere.
The second deficiency consists in the necessity of adjusting the
fitting parameter $\eta$, as different $\eta$ values are required
ad hoc to match observed mass-loss rates from different types of
giants and supergiants.  The same is true, 
if reasonable mass loss yields and final masses are to be achieved through
stellar evolution models with prescribed mass loss. For more evolved 
AGB giants, the Reimers relation is better replaced by, e.g., \cite{dej88}, 
which suggests up to three times as much mass-loss for the tip-AGB
\citep{schr01}.  More recently, the Reimers relation also
fails to describe revised mass loss rates from K and M giant stars,
based on updated Ca~II ionization balances which consider 
photoionization radiation deduced from FUSE spectra \citep{har04}. 
For further updated information on mass-loss mechanisms see, e.g.,
the review by \cite{wil00}.

In the present work, we overcome these deficiencies by adopting a
more physical picture.  In our approach, the non-radiative energy input
into the wind is assumed to be given by the turbulent energy density,
within the chromosphere or underneath, possibly related to the
manifestation of (magneto-)acoustic waves.  This approach appears to be
consistent with the major conclusion by \cite{jud91}, who presented a
detailed empirical analysis of the global thermodynamical properties of
the outer atmospheres and winds of a set of well-studied cool giant and
supergiant stars.  They concluded that ``[...] mass loss rates are not
strongly dependent on the actual physical processes driving the winds
[suggesting] that nonlinear processes act to regulate wind energy
fluxes". Furthermore, we assume that the mass loss rate
depends on the characteristic chromospheric height, which dictates the
amount of energy required to lift the wind out of the potential well of the
star.  This simplified model results in a mass loss formula akin to that
by \cite{rei75,rei77}.  However, it contains two additional factors, one
depending on the effective temperature and the second on the surface
gravity of the star.  This improves the agreement with observed
mass loss rates for different types of stars, without the need to adjust
the fitting parameter.  In \S 2, we describe our theoretical approach.
In \S 3, we discuss tests and applications of our new formula, and in \S 4,
we give our conclusions. 

%%%%%%%%%%%%%%%%%%%%%%%%%%%%%%%%%%%%%%%%%%%%%%%%%%%%%%%%%%%%%%%%%%%%%%%%%%%%

\section{Theoretical Approach}

In our simplified model, we regard the wind to result from a spill-over
of the extended, highly turbulent giant chromosphere and its reservoir
of mechanical energy, possibly associated with waves.  Even though no
details of any theoretical models are considered, the following derivation
will be guided by the assumption of Alfv\'en waves, owing to their success
in describing stellar winds as obtained for $\alpha$~Boo (K1.5~III)
\citep{har80}, $\zeta$~Aur~A (K4~Ib) \citep{kui89}, and $\alpha$~Ori (M2~Iab)
\citep{har84,air00}.  The relevant mechanical energy flux $F_{\rm M}$ may
thus be due to magnetic energy generation, a consequence of non-isotropic
chromospheric turbulence, or a combination of both.

Turbulence is a well-known feature of stellar photospheres
\cite[e.g.,][and references therein, among more recent literature]{gra92},
and of cool star chromospheres.  E.g., \cite{car96}
obtained empirical constraints on the chromospheric macroturbulence and
flow velocities for various K and M (super-)giants based on C~II] from
HST-GHRS spectra, ranging from 24~km~s$^{-1}$ ($\alpha$~Tau; K5~III) to
35~km~s$^{-1}$ ($\alpha$~Ori; M2~Iab), which are in principle sufficient
to overcome the gravitational potential of the star.  Nevertheless, the
chromospheric turbulent energy density relevant for the generation of winds
is not exactly known, largely because of the difficulty
of distinguishing between isotropic and non-isotropic turbulence.
Examples of wave-driven wind models have also been given.  For instance,
\cite{air00} proposed a time-dependent, 2.5-D Alfv\'en wave wind model,
resulting in a time-averaged mass loss rate commensurate with the recent
semi-empirical chromosphere and wind model by \cite*{har01} based on
NRAO VLA radio data.

Aside from the amount of utilized mechanical energy flux,
the stellar mass loss rate
is also expected to depend on the characteristic chromospheric radius
$R_{\rm Chr}$, which dictates the amount of wind energy 
$dE_{\rm Wind}$ needed by a mass element $dM = \dot{M} dt$ 
to overcome the gravitational potential of the star.  For the wind
energy balance, we thus obtain
\begin{equation}
dE_{\rm Wind} \ \simeq \ \frac{G M_* \dot{M} dt}{R_{\rm Chr}}
              \ \propto \ F_{\rm M} \cdot 4\pi R_*^2 dt 
\end{equation}
with $\dot{M}$ as mass loss rate, and $R_*$, $M_*$ as stellar radius and mass,
respectively, $F_{\rm M}$ as mechanical energy flux, and $G$ as 
gravitational constant\footnote{Historically, a similar approach
has been undertaken by \cite{fus75}, who assumed that cool star winds are
due to acoustic waves.  Later on, this assumption has been invalidated by
detailed model simulations, which showed that acoustic waves fail to
support significant mass loss as they dissipate most of their energy
immediately beyond the stellar photosphere, and are thus incapable of
meeting the potential energy requirement of stellar winds
\citep[e.g.,][and references therein]{cun90,sut95}.}.

A large body of literature has been devoted to describing convective
turbulence of stellar atmospheres and the generation of waves
as function of the fundamental stellar parameters.
\cite{ste81} studied the generation of waves by
turbulent motions in stellar atmospheres largely based on analytic
means.  He found that the acoustic wave energy flux is given as
$F_{\rm M} \propto T_{\rm eff}^{6.1}$ (monopole term),
$T_{\rm eff}^{10.4}$ (dipole term), and $T_{\rm eff}^{14.6}$ (quadrupole
term)(see representation by Ulmschneider [1989]),
noting that the monopole term is most closely and the quadrupole term is
least closely related to mass loss generation. 
Models by \cite{boh84} deduce as temperature dependence for the
monopole, dipole, and quadrupole terms, $T_{\rm eff}^{4.15}$,
$T_{\rm eff}^{8.78}$, and $T_{\rm eff}^{13.85}$, respectively, and for the
combination of those terms $T_{\rm eff}^{9.75}$.  Analytic models for
magnetic wave generation reveal a temperature dependence of
$T_{\rm eff}^{7.5}$ for the Alfv\'en mode and  
$T_{\rm eff}^{8}$ for the magnetic modes combined \citep{mus88}.

More recent work has resulted in vast improvements of these models,
particularly with respect to the models for the stellar convection
zones and the description of the turbulent frequency spectra ---
see, e.g., \cite{mus04} for a recent review on those results.
Unfortunately, the authors usually refrain from giving a fitting
formula for the dependence of the wave energy fluxes on the
stellar effective temperatures, supposedly because of the large number
of free parameters, particularly in magnetic models.  Nevertheless,
based on the fact that the solar wind \citep[e.g.,][]{ong97} and
massive stellar winds \citep[e.g.,][]{ros95,air00}
are likely to be accelerated by the momentum deposition by Alfv\'en
waves, we consider a dependence as $T_{\rm eff}^{7.5}$ to be
most representative.  There is also another feature inherent to
Alfv\'en waves motivating this choice: they are known to be
essentially non-dissipative in the lower and middle chromosphere
--- see discussion by \cite{cha95} and \cite{boy96}, pointing to
the prevalence of low-amplitude waves in those stars ---
implying that the same $T_{\rm eff}$-exponent holds in the region
of wave generation as well as where the onset of mass loss occurs.
Incidently, the same exponent is also found as temperature dependence
of Mg~II {\it h}, {\it k} emission in stars (both dwarfs and giants)
of minimal activity, a likely indicator of the overall chromospheric
energy density \cite*[][see their Fig.~15]{buc98}.

If the mechanical energy flux $F_{\rm M}$ utilized for generating stellar
mass loss is assumed as $F_{\rm M} \propto T_{\rm eff}^{7.5}$, the
surface-integrated mechanical energy flux $L_{\rm M}$ can now be expressed as
\begin{equation}
L_{\rm M} \ = \ F_{\rm M} \cdot 4 \pi R_*^2  \ \propto \
F_{\rm M} \cdot L_* / T_{\rm eff}^4 \ \propto \ L_* \cdot T_{\rm eff}^{3.5}
 \ \ \ .
\end{equation}
Next we consider the characteristic chromospheric radius $R_{\rm Chr}$.
For cool giants and supergiants, no well-defined boundary between the
chromosphere and the wind exists.  Hence, we use the sonic point of the
average velocity field as reference.  For the well-studied K supergiant
$\zeta$~Aur (with $\log{g_*} \simeq 0.8$), $R_{\rm Chr}$ is found to be
close to $2 R_*$ \citep{baa96}, and for general giants and supergiants,
$(R_{\rm Chr}-R_*)/R_*$ is assumed to vary as $g_*^{-1}$, which gives 
\begin{equation}
R_{\rm Chr} \ = \ R_* \cdot \Bigl(1 + \frac{g_{\odot}}{4300 \cdot g_*}\Bigr)
 \ \ \ .
\end{equation}
With the above temperature dependence of the mechanical energy flux 
(eq.~[2]) and chromospheric radius $R_{\rm Chr}$ (eq.~[3]), we finally
obtain as mass loss rate $\dot{M}$, see eq.~(1),
\begin{equation}
\dot{M} \ = \ \eta \cdot \frac{L_* R_*}{M_*} \cdot 
{\Bigl(\frac{T_{\rm eff}}{4000~{\rm K}}\Bigr)}^{3.5} \cdot
\Bigl(1+\frac{g_{\odot}}{4300 \cdot g_*}\Bigr)
\end{equation}
with $R_*$, $M_*$, and $L_*$ as stellar radius, mass, and luminosity
given in solar units, and $g_*$ and $g_\odot$ as stellar and solar
surface gravity, respectively.  This is, apart from the two new factors, 
indeed the old Reimers law.  To satisfy the well-constrained RGB
mass-loss of globular cluster stars (see \S~3), the fitting parameter
$\eta$ will be set to $8 (\pm1) \times 10^{-14} M_{\odot}$~yr$^{-1}$.

%%%%%%%%%%%%%%%%%%%%%%%%%%%%%%%%%%%%%%%%%%%%%%%%%%%%%%%%%%%%%%%%%%%%%%%%%%%%

\section{Tests and Applications}

For the newly developed mass loss formula various tests and applications
have been devised.  In particular, we want to obtain insight into the
importance of the new factors given by the stellar effective temperature
$T_{\rm eff}$ and gravity $g_*$.  In fact, for ordinary giants the two new
factors do not make much difference, which explains the long-lasting success
of the Reimers relation.  In particular, the $T_{\rm eff}^{3.5}$ factor is,
despite its high power, restricted in its impact by the small band 
of relevant effective temperatures (3000 to 4500~K), and the $g$-sensitive
factor remains of the order of one for all but the smallest gravities.
In fact, as previously discussed, the $T_{\rm eff}$ exponent in eq.~(4)
is somewhat uncertain.  However, due to the narrow band of relevant
effective temperatures, the overall results would still stand
if $T_{\rm eff}^3$ or $T_{\rm eff}^4$ were used instead.

Nevertheless, we have obtained evidence for the extra dependence on 
$T_{\rm eff}$ from a comparative study of the RGB mass loss of globular 
cluster (GC) stars with very different metallicity and, accordingly, different 
effective temperatures on their RGBs.  The mass lost on the RGB of a GC 
is very well constrained by modelling the stars at the blue end of the 
Horizontal Branch (HB), for which the HRD position is very mass sensitive. 
The remaining uncertainty is about 15 \% in absolute terms and
much better in relative terms.  In fact, the long time (on a dynamic 
time-scale) spent on the RGB effectively evens out most of the inherent 
variability of these stellar winds.  This is a big advantage
over the kind of snapshots obtainable from observing individual winds
directly.  A residual star-to-star variation of $<20$\%, on the other hand,
is sufficient to explain the full spread of an HB.

As test cases, we consider two globular clusters, which are NGC~5904 and
NGC~5927.  NGC~5904 has a significant metal underabundance of [Fe/H] = -1.29
(normal for globular clusters, Z=0.001, see Fig.~1), whereas NGC~5927 only
has a marginal underabundance of [Fe/H] = -0.37 (near Z=0.01, see Fig.~2).
We use the photometric data and metallicities provided by \cite{pio02} and
have plotted our evolution tracks directly into their cluster CMDs.
Clearly, the Reimers relation cannot
reproduce both cases with the same $\eta_{\rm R}$, while our new relation can!
In particular, the extreme blue end of the Horizontal Branch (HB) of 
NGC~5904 (Fig.~1, {\it bottom panel}) demands HB stellar masses $M_{\rm HB}
\simeq 0.60 M_{\odot}$, with a He-core mass of $M_c \simeq 0.49 M_{\odot}$.  With
an approximate initial mass of $M_i \simeq 0.86 M_{\odot}$, consistent
with an age of about 12 billion years, this corresponds to a total 
RGB mass loss of 0.26 $M_{\odot}$ for the individual GC stars.  With
our new mass loss relation with $\eta = 0.8 \times 10^{-13}$, our evolution
models achieve exactly this HB mass.  To get the same result with the
old Reimers law, we would need a $\eta_{\rm R} = 2.4 \times 10^{-13}$.
On the other hand, the small extent of the HB of 
NGC~5927 (Fig.~2, {\it bottom panel}) demands HB stellar masses $M_{\rm HB}
\simeq 0.71 M_{\odot}$, possibly slightly more, but certainly not less, 
with a He-core mass of $M_c \simeq 0.48 M_{\odot}$.  With
$M_i \simeq 0.99 M_{\odot}$, consistent with an age of about 11 billion yrs,
this corresponds to a total RGB mass loss of 0.28 $M_{\odot}$ --- exactly
as achieved by our evolution models with $\eta = 0.8 \times 10^{-13}$ and
our new mass-loss relation.  To get the same HB mass with the Reimers relation,
we would need $\eta_{\rm R} = 2.0 \times 10^{-13}$.  On the other hand, a value of
$\eta_{\rm R}$ of $2.4 \times 10^{-13}$ can clearly be ruled out: it would produce
HB stars with $M_{\rm HB} \simeq 0.63 M_{\odot}$, which are already far too blue.

The role of the new, gravity-related factor appearing in our new mass loss
relation can be explored by assessing luminous low-gravity stars, which play a very
important role in the overall mass loss during the stellar lifetime.  A good
test candidate is the well-studied star $\alpha$~Ori, which has some
circumstellar dust but is still below its critical luminosity for
possessing a truly dust-driven wind.  The mass loss rate was observed as 
$\dot{M}_{\rm obs} = 3.1 (\pm 1.3) \times 10^{-6} M_{\odot}$~yr$^{-1}$
\citep{har01}, assuming a distance of $d=131$~pc ($\pm 25$\%) according to
the Hipparcos parallax.  The corresponding luminosity is
$L_* = 5.4 \times 10^4 L_{\odot}$, and with an angular diameter of 56~mas,
$T_{\rm eff}$ is given as 3140~K, for which matching evolution tracks imply
a mass of $M_* = 10 M_{\odot}$ ($\pm 30$\% depending on $L_*$).  For these
parameters, our new mass loss relation (with $\eta = 0.8 \times 10^{-13}$) 
yields $2.2 \times 10^{-6} M_{\odot}$~yr$^{-1}$, 
while the Reimers law gives only $0.8 \times 10^{-6} M_{\odot}$~yr$^{-1}$ 
(with $\eta_{\rm R} = 2 \times 10^{-13}$).
The distance uncertainty for $\alpha$~Ori affects
all three mass loss rates (including $\dot{M}_{\rm obs} 
\propto d^2$), but not so much their ratios.  For example, 
if $d$ was 25\% larger and $M_* \simeq 13 M_{\odot}$, then 
$\dot{M}_{\rm obs} = 4.8 \times 10^{-6} M_{\odot}$~yr$^{-1}$, 
our value would be $4.0 \times 10^{-6} M_{\odot}$~yr$^{-1}$,
whereas the Reimers law would give 
$1.3 \times 10^{-6} M_{\odot}$~yr$^{-1}$.
In any case, for such low gravities ($\log g \simeq -0.35$), 
$R_{\rm Chr}/R_*$ increases significantly, resulting in an extra boost
of mass loss.

%%%%%%%%%%%%%%%%%%%%%%%%%%%%%%%%%%%%%%%%%%%%%%%%%%%%%%%%%%%%%%%%%%%%%%%%%%%%

\section{Conclusions}

We derived a new semi-empirical relation for the mass loss of cool winds,
which so far has frequently been described by Reimers'
law.  Physically, the Reimers relation appears to suggest a picture
in which the wind material is lifted from the photosphere by using a 
certain fraction of the stellar luminosity --- even though it is well known
that, with the exception of molecule-driven and dust-driven winds, cool
star winds are not related to any type of radiation pressure.  This apparent
contradiction has now been resolved.

The new relation is based on theoretical arguments assuming that the wind
results from the turbulent energy density, within the chromosphere or
underneath, possibly related to the manifestation of magnetoacoustic
waves as, e.g., Alfv\'en waves.  Furthermore, the mass loss rate is assumed
to depend on the chromospheric extent, which dictates the amount of energy
required to lift the wind out of the potential well of the star.
A more detailed analysis shows that the new mass loss formula is
not applicable to molecule-driven, dust-driven, and pulsational winds,
as in those cases highly temperature-sensitive feed-back mechanisms exist,
resulting in a steeper dependence of the mass loss rate on the stellar
parameters, including the metallicity, which is not reflected by the new formula.
Moreover, pulsational winds are more episodic in nature,
whereas the new mass loss formula only descibes time-averaged mass loss
behavior.  Also note that the new mass-loss formula is not valid for
stars like the Sun, where information exists that different types of
mass loss processes, resulting in slow and fast wind, operate on
different horizontal and vertical scales, which is beyond the theoretical
framework of this paper.

The new relationship mostly reproduces
the original Reimers law, except that it includes two additional factors,
which further improve the agreement with observed mass loss rates,
especially for (super-)giants with very low gravity.
This improved agreement can be interpreted as an indirect
validation of Alfv\'en waves as being primarily responsible for mass loss
generation in those stars.  A highly sensitive
calibration of the new relation's fitting factor $\eta$ has been achieved
by modelling the mass lost on the RGB by stars near the blue end of the
Horizontal Branch, using two globular clusters of very different metallicity 
(NGC~5904, NGC~5927). 
Further studies, considering sets of well-studied K and M-type giant and
supergiant stars, including comparisons with the various mass loss
formulas from the literature, will be given in the near future.

%%%%%%%%%%%%%%%%%%%%%%%%%%%%%%%%%%%%%%%%%%%%%%%%%%%%%%%%%%%%%%%%%%%%%%%%%%%%

%% Included in this acknowledgments section are examples of the
%% AASTeX hypertext markup commands. Use \url without the optional [HREF]
%% argument when you want to print the url directly in the text. Otherwise,
%% use either \url or \anchor, with the HREF as the first argument and the
%% text to be printed in the second.

\acknowledgments

\bigskip
\noindent
We are grateful to Z.~E. Musielak, D. Sch\"onberner, R.~C. Smith, and
B.~E.~J. Pagel for helpful discussions.  The paper also benefited from
comments by three anonymous referees.
This work has been supported by NSF under grant ATM-0087184 (M.~C.).




%%clearpage to be removed

%%clearpage to be removed
\clearpage

%  \noindent
%  {\bf Figure Caption}
%
%
%%%  Figure 1
%
  \begin{figure*}
  \includegraphics[angle=270,width=12cm]{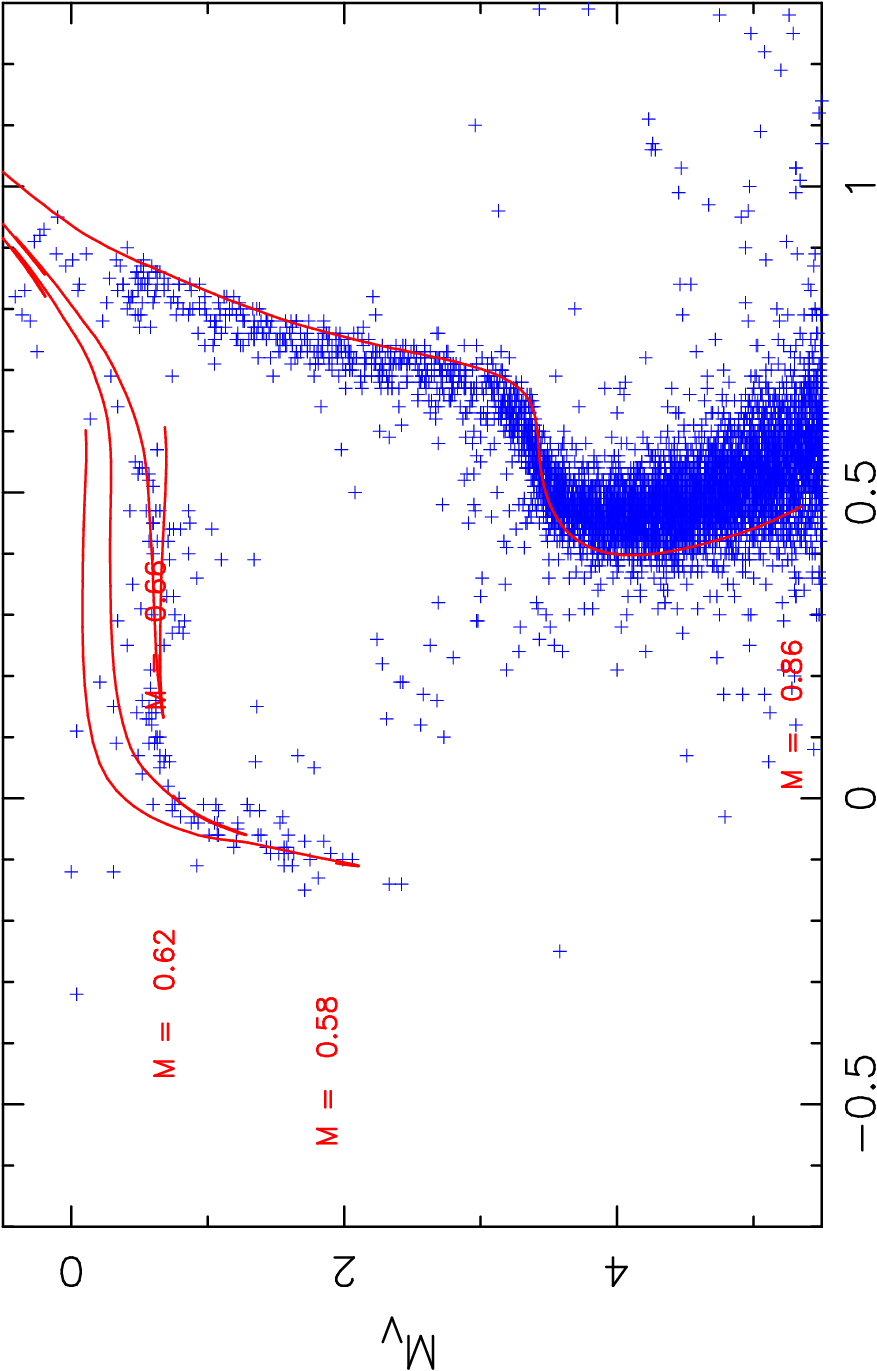}
  \vspace{0.4in}
  \includegraphics[angle=270,width=12.04cm]{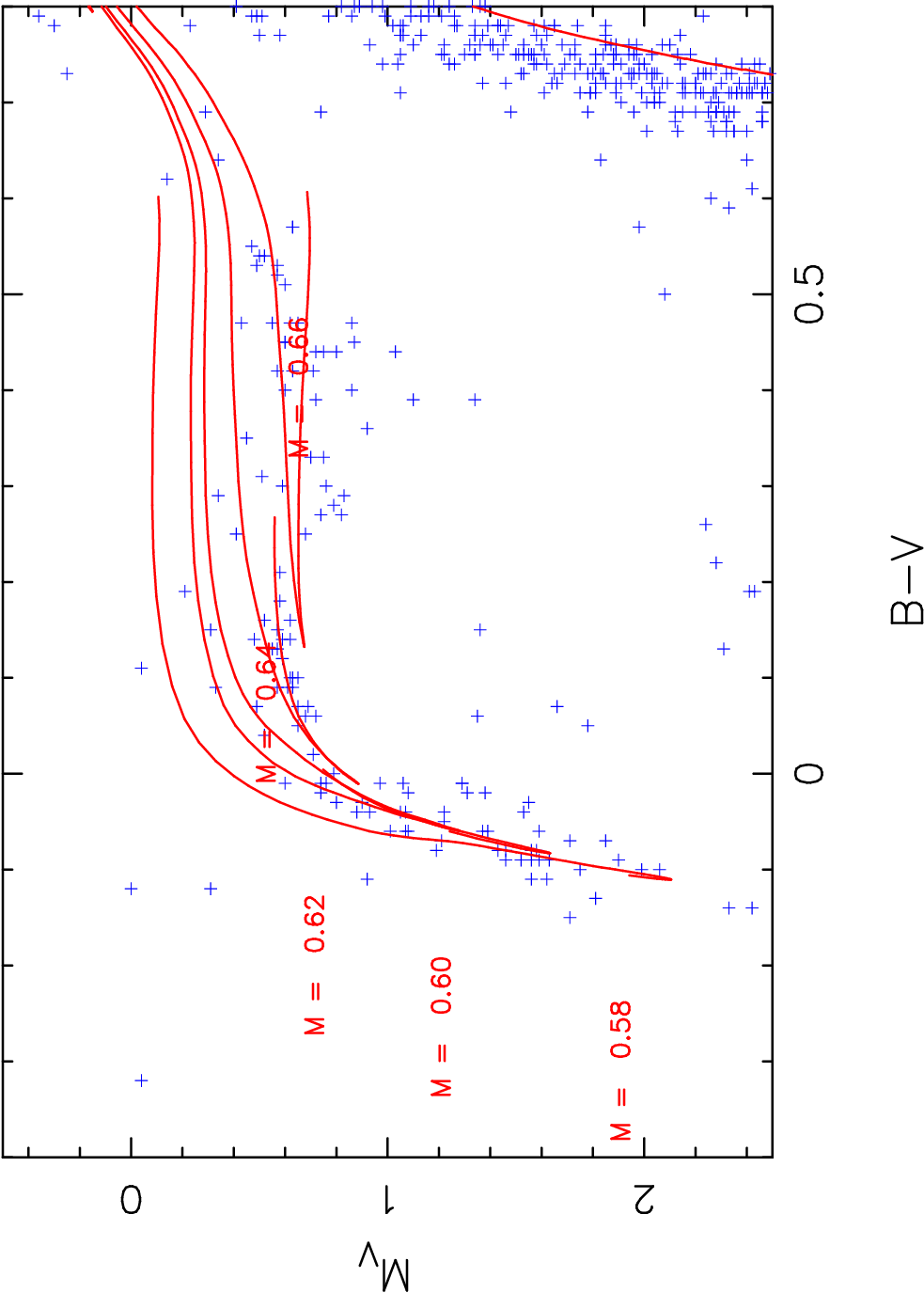}
  \caption{{\it Top panel:} Color-Magnitude Diagram of the globular
   cluster NGC~5904.   {\it Bottom panel:}
   Position of the Horizontal Branch with Z=0.001.}
  \end{figure*}
%
%
%%%  Figure 2
%
  \begin{figure*}
  \includegraphics[angle=270,width=12cm]{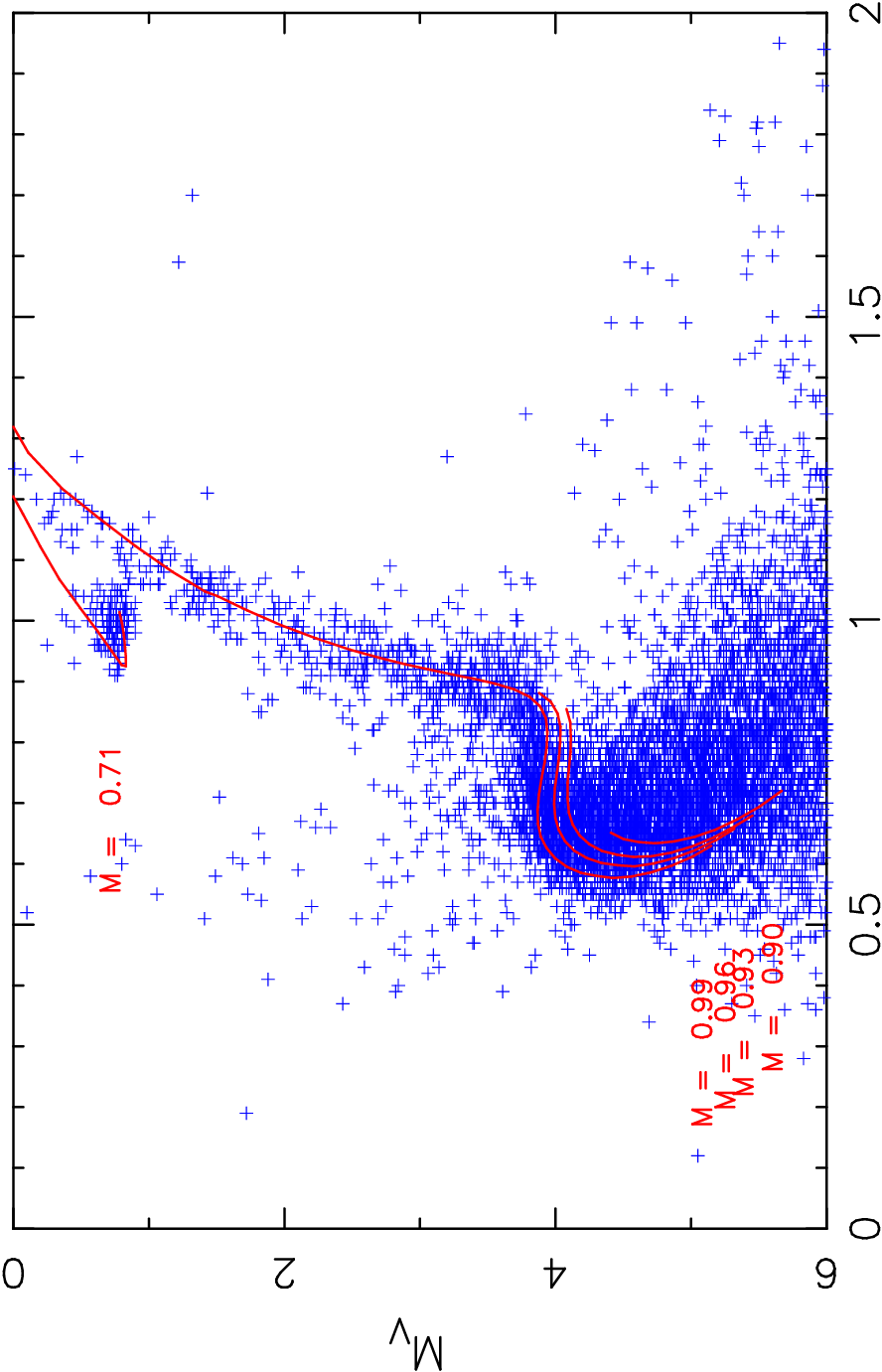}
  \vspace{0.4in}
  \includegraphics[angle=270,width=12.17cm]{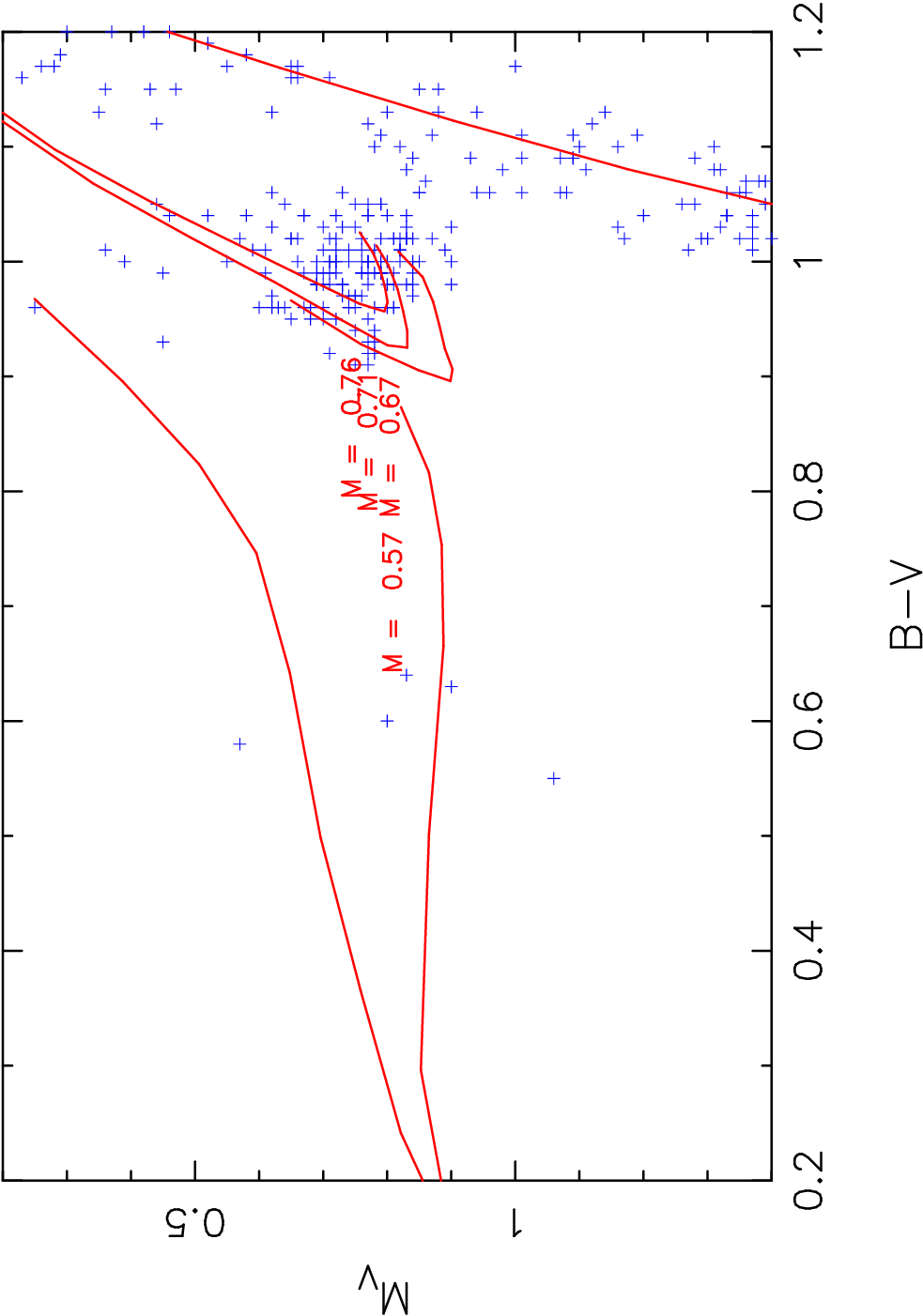}
  \caption{{\it Top panel:} Color-Magnitude Diagram of the globular
   cluster NGC~5927.  {\it Bottom panel:}
   Position of the Horizontal Branch with Z=0.009.}
  \end{figure*}
%
%
%  %%clearpage to be removed
%  \clearpage



\end{document}